# The loss of interest for the euro in Romania

Claudiu Tiberiu ALBULESCU[1*] and Dominique PÉPIN[2]

[1] Management Department, Politehnica University of Timisoara

[2] CRIEF, University of Poitiers

**Abstract**

We generalize a money demand micro-founded model to explain Romanians' recent loss of interest for the euro. We show that the reason behind this loss of interest is a severe decline in the relative degree of the euro liquidity against that of the Romanian leu.

---

[*] Corresponding author. E-mail addresses: claudiu.albulescu@upt.ro, claudiual@yahoo.com.

# 1. Introduction

Romania joined the European Union (EU) in 2007 being now one of the Euro area candidate countries. Long before Romania's entrance to the EU, the leu and the euro went hand in hand as the main transactions and savings currencies. However, since September 2001, the euro holding has considerably diminished as compared to that of the domestic currency.

Let $M_t$ denote the Romanian domestic money holding, while $M_t^*$ is the euro holding. Assuming that $S_t$ is the exchange rate, then $S_t M_t^*$ represents the euro holding denominated in lei. Figure 1 shows the drop of the $S_t M_t^* / M_t$ ratio over the period 2001:M9-2015:M11.

Figure 1. Romanians' euro holding to domestic money holding ratio

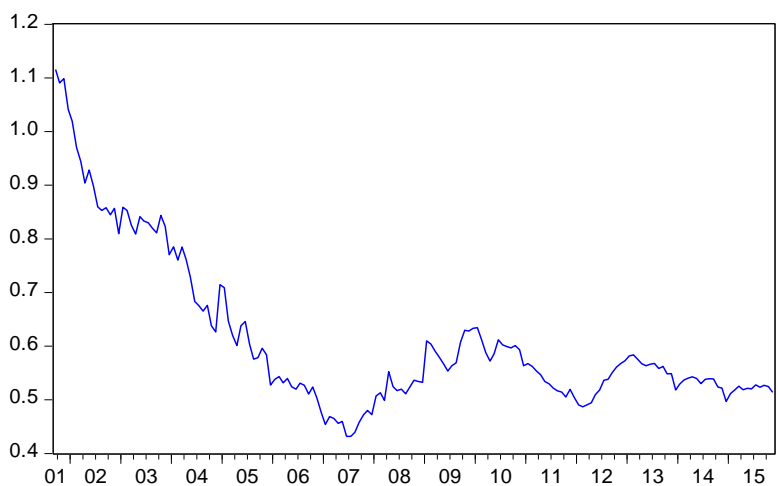

*Source: Own computations based on monthly bulletins of the National Bank of Romania*

The literature on money demand provides no explanation for such a trend. The money demand in the CEE countries is empirically investigated by Van Aarle and Budina (1996), Mulligan and Nijsse (2001), Dreger et al. (2007), Fidrmuc (2009), or Dritsaki and Dritsaki (2012). Single-country analyses of money demand are conducted *inter-alia* by Komárek and Melecký (2004) for the Czech Republic, by Siliverstovs (2008) for Latvia, or by Hsieh and Hsing (2009) for Hungary. The money demand in Romania was investigated by Andronescu et al. (2004) and Ruxanda and Muraru (2011). None of these papers provides a micro-founded theoretical model to justify the specification of their empirical money demand functions. Albulescu and Pépin (2016) represent an exception.

Our contribution to the existing literature is twofold. First, we generalize the micro-founded model of Albulescu and Pépin (2016) by assuming that the relative liquidity degree of the euro against that of the leu is changing. Second, we apply the new model on the Romanian case and explain the loss of interest for the euro during the last period.

**2. A money demand model in an open economy**

Generalizing Albulescu and Pépin (2016), we suppose that the lifetime utility function of the domestic agent is:

$$V_t = E_t \left[ \sum_{i=0}^{\infty} \beta^i U\left( \frac{X_{t+i}}{P_{t+i}}, \frac{M_{t+i}}{P_{t+i}}, \frac{S_{t+i}M^*_{t+i}}{P_{t+i}}; \varepsilon_{t+i}, t+i \right) \right], \qquad (1)$$

where $X_t$ is the monetary consumption spending denominated in lei, $P_t$ is the price index, $\varepsilon_t$ is a stationary stochastic process and t is a deterministic trend. $E_t[.]$ is the expectation conditional upon the information available at time t and the presence of $\varepsilon_t$ and of trend t in the utility function indicates that its properties are subject to changes. This utility specification is based on the assumption that the representative agent holds foreign and domestic money in relation with his total consumption, with no distinction between his consumption of foreign and domestic goods.

Now suppose that the utility function takes the form:

$$U\left( \frac{X_t}{P_t}, \frac{M_t}{P_t}, \frac{S_t M^*_t}{P_t}; t, \varepsilon_t \right) = \left( \frac{X_t}{P_t} \right)^{1-\theta} \left\{ \delta(t,\varepsilon_t)\left( \frac{M_t}{P_t} \right)^{\gamma} + (1-\delta(t,\varepsilon_t))\left( \frac{S_t M^*_t}{P_t} \right)^{\gamma} \right\}^{\frac{\theta}{\gamma}}, \quad \gamma = \frac{\sigma-1}{\sigma}, \qquad (2)$$

where $\sigma = 1/(1-\gamma)$ is the elasticity of substitution between the leu and the euro and $\delta(t,\varepsilon_t)$ is a function of t and $\varepsilon_t$ (the share parameter).

If the elasticity $\sigma$ is high, it is easier to replace one currency by another, which represents a proof of monetary integration (Fidrmuc, 2009). Therefore, if $\sigma > 1$ we have substitutability between currencies, while a value $\sigma < 1$ indicates their complementarity. In their simplified model, where $\delta(t,\varepsilon_t) = \delta \ \forall t$, Albulescu and Pépin (2016) find that the elasticity of substitution between the leu and the euro is weak, ranging between 0.3 and 0.5 under different estimations, and reject thus the hypothesis of monetary integration with the Euro area.

The expression $\left\{ \delta(t,\varepsilon_t)\left( \frac{M_t}{P_t} \right)^{\gamma} + (1-\delta(t,\varepsilon_t))\left( \frac{S_t M^*_t}{P_t} \right)^{\gamma} \right\}^{\frac{1}{\gamma}}$ is the liquidity production function and the term $(1-\delta(t,\varepsilon_t))/\delta(t,\varepsilon_t)$ measures the liquidity degree of the euro against the leu in the eyes of the Romanian representative agent. It indicates the amount of euros needed to produce the same liquidity service as one leu. The liquidity service provided by a currency is likely to be influenced by any institutional change, by financial innovations or technological developments in payment systems, or, in this particular case, by blockages in the real estate market, hence the hypothesis of a changing degree of liquidity.

The real consumption and the CES liquidity production function are next combined according to a Cobb-Douglas utility function with a consumption elasticity of $1-\theta$ and a liquidity elasticity of $\theta$, while the parameters are restricted so that $0<\theta<1$, $0<\delta(t,\varepsilon_t)<1$ and $0 \leq \sigma < +\infty$.

The representative agent faces the budget constraint:

$$M_{t-1}(1-\phi) + S_t M^*_{t-1}(1-\phi) + B_{t-1}(1+i_t) + S_t B^*_{t-1}(1+i^*_t) + Z_t = X_t + M_t + S_t M^*_t + B_t + S_t B^*_t, \quad (3)$$

where $B_t$ is the value (in lei) of domestic bond holding, $B^*_t$ is the value (in euro) of foreign bond holding, $Z_t$ are the non-financial incomes and monetary transfers from the government, and $i_{t+1}$ and $i^*_{t+1}$ are the nominal domestic and foreign interest rates. Because bonds are nominally risk-free, $i_{t+1}$ and $i^*_{t+1}$ are known at time t.

The parameter $\phi$ represents the proportional cost the agent faces for holding money (for simplification this cost is considered the same for cash and deposits and is fixed as in Lucas and Nicolini (2015) at 1% per year, that is 0.082953% on a monthly basis). It stands for the charges related to the use of money (cost of a bank card, cash theft or loss…).

Consider now the optimization problem of the representative agent who maximizes (1) with respect to $\frac{M_t}{P_t}, \frac{S_t M^*_t}{P_t}, \frac{B_t}{P_t}$ and $\frac{S_t B^*_t}{P_t}$, under (2) and (3). If we consider that $\delta(t,\varepsilon_t)$ is perfectly observable at date t, the first-order conditions $E_t[\partial V_t/\partial(M_t/P_t)]=0$ and $E_t[\partial V_t/\partial(B_t/P_t)]=0$ from Albulescu and Pépin (2016) lead to:

$$\frac{\theta}{1-\theta}\left(\delta(t,\varepsilon_t)\left(\frac{M_t}{P_t}\right)^\gamma + (1-\delta(t,\varepsilon_t))\left(\frac{S_t M^*_t}{P_t}\right)^\gamma\right)^{-1}\left(\frac{X_t}{P_t}\right) = \frac{oc_{t+1}}{\delta(t,\varepsilon_t)}\left(\frac{M_t}{P_t}\right)^{1-\gamma}, \quad (4)$$

where $oc_{t+1} = \frac{i_{t+1}+\phi}{1+i_{t+1}}$ is the opportunity cost of domestic money holding, while the remaining two first-order conditions, $E_t[\partial V_t/\partial(S_t M^*_t/P_t)]=0$ and $E_t[\partial V_t/\partial(S_t B^*_t/P_t)]=0$ allow to obtain:

$$\frac{\theta}{1-\theta}\left(\delta(t,\varepsilon_t)\left(\frac{M_t}{P_t}\right)^\gamma + (1-\delta(t,\varepsilon_t))\left(\frac{S_t M^*_t}{P_t}\right)^\gamma\right)^{-1}\left(\frac{X_t}{P_t}\right) = \frac{oc^*_{t+1}}{1-\delta(t,\varepsilon_t)}\left(\frac{S_t M^*_t}{P_t}\right)^{1-\gamma}, \quad (5)$$

where $oc^*_{t+1} = \dfrac{i^*_{t+1} + \phi}{1 + i^*_{t+1}}$ is the opportunity cost of euro holding.

As the left-hand terms of equations (4) and (5) are the same, it comes:

$$\ln\left(\frac{S_t M^*_t}{M_t}\right) = \sigma \ln\left(\frac{1 - \delta(t, \varepsilon_t)}{\delta(t, \varepsilon_t)}\right) + \sigma\left(\ln oc_{t+1} - \ln oc^*_{t+1}\right). \qquad (6)$$

Equation (6) describes the euro holding relative to the leu holding. If the opportunity cost of holding euros increases more than the opportunity cost of holding lei, the Romanians abandon the euro in favor of the leu, and this is what really happened. This substitution effect is even stronger if the elasticity of substitution between currencies is high. Figure 2 shows that the opportunity cost spread $\ln oc_{t+1} - \ln oc^*_{t+1}$ decreases since 2001, even if in an irregular way, given the impact of the 2008 financial crisis on the interest rate spread.

Figure 2. The opportunity cost spread of holding domestic money against euro

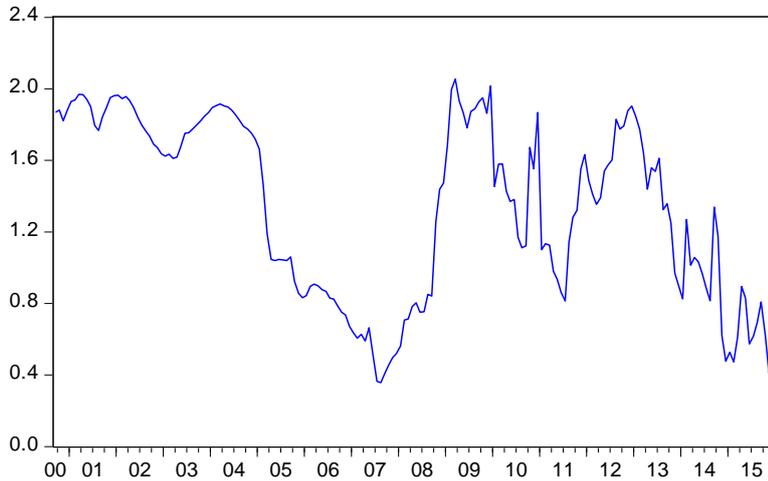

*Source: Own computations based on Albulescu and Pépin (2016)*

Over the timespan 2001:M9 to 2015:M9, the linear correlation coefficient between $\ln(S_t M^*_t / M_t)$ and $\ln oc_{t+1} - \ln oc^*_{t+1}$ is 0.341. The coefficient's positive value shows that part of the euro holding decline undoubtedly relates to the downfall of the Romanian short interest rate, as compared to the Euro area interest rate. In the long-run, $\ln(S_t M^*_t / M_t)$ should be perfectly explained by $\ln oc_{t+1} - \ln oc^*_{t+1}$ if $\delta(t, \varepsilon_t) = \delta \; \forall t$. However, this is not the case because the correlation coefficient is considerably smaller than 1. Consequently, other factors might explain this strong loss of interest for the euro, like a changing liquidity degree of the euro.

Indeed, equation (6) also shows that the euro demand depends on $(1-\delta(t,\varepsilon_t))/\delta(t,\varepsilon_t)$ ratio. A decrease in this ratio is associated with a decrease in the liquidity service provided by the euro and thus a reduction in the demand for euro. It remains to assess the relevance of this explanation by estimating the function $\delta(t,\varepsilon_t)$.

If $\ln(M_t/S_tM_t^*)$ and $(\ln oc_{t+1} - \ln oc_{t+1}^*)$ are I(1), a simple way for $\delta(t,\varepsilon_t)$ parametrization is to suppose the existence of three $\upsilon_0, \upsilon_1$ and $\upsilon_2$ parameters such as:

$$\sigma \ln\left(\frac{1-\delta(t,\varepsilon_t)}{\delta(t,\varepsilon_t)}\right) = \upsilon_0 + \upsilon_1 t + \upsilon_2 t^2 + \varepsilon_t. \quad (7)$$

Inserting (7) in (6), we obtain the cointegration equation which allows the estimation of the elasticity of substitution $\sigma$ and the ratio $(1-\delta(t,\varepsilon_t))/\delta(t,\varepsilon_t)$:

$$\ln\left(\frac{S_t M_t^*}{M_t}\right) = \upsilon_0 + \upsilon_1 t + \upsilon_2 t^2 + \sigma(\ln oc_{t+1} - \ln oc_{t+1}^*) + \varepsilon_t. \quad (8)$$

## 3. Empirical evidence

We use monthly statistics from 2001M9 to 2015M11 to estimate equation (8). The money market rate is used to compute the discounted interest rate (data come from IMF International Financial Statistics). The structure of bank deposits is used as a proxy of the money demand structure (data are extracted from the National Bank of Romania monthly bulletins). We apply the Hansen's (1992) test to check the existence of a long-run relationship and we estimate the cointegration equation (8) by the Fully-Method Ordinary Least Squares (FMOLS) method.

Before estimating equation (8), we verify if our series are I(1). The application of the ADF and PP unit root tests prove the existence of a unit root for both processes (Table 1).

Table 1. Unit root tests

| Variables | Test ADF (with intercept) | Test ADF (with trend and intercept) | Test PP (with intercept) | Test PP (with trend and intercept) |
|---|---|---|---|---|
| $\ln(S_t M_t^*/M_t)$ | -2.89* | -2.26*** | -3.01* | -2.22*** |
| $\ln oc_{t+1} - \ln oc_{t+1}^*$ | -1.74*** | -2.00*** | -1.74*** | -1.99*** |

*Notes: (i) the null hypothesis is the presence of unit root; (ii) \*, \*\*, \*\*\* means a p-value for the t-statistic >1%, >5% and >10% respectively. Otherwise said, \*\* implies that null hypothesis of unit root cannot be rejected at 5% significance level.*

The results of the FMOLS estimation for equation (8) are highlighted in Table 2. The Hansen's test accepts the existence of a cointegrating relationship at 10%, result that validates the theoretical specification of our model.

Table 2. FMOLS estimation

| Parameter | $\upsilon_0$ | $\upsilon_1$ | $\upsilon_2$ | $\sigma$ |
|---|---|---|---|---|
| Estimated value | -0.037619 | -0.012215*** | 0.000042*** | 0.201694*** |
| t-statistic | -0.49 | -10.59 | 8.97 | 8.88 |
| Probability | 0.6221 | 0.0000 | 0.0000 | 0.0000 |
| $R^2 = 0.91$ | Lc statistic = 0.561928 | | Probability Lc test = 0.1358 | |

*Notes: (i) \*\*\*, \*\*, \* means significance at 1%, 5% et 10% significance level; (ii) Probability means p-value of the test of nullity ; (iii) Lc statistic is the statistic of the Hansen's cointegration test (Hansen, 1992) and Probability Lc test is the p-value of test (the null hypothesis is cointegration).*

All the parameters are significant and the $\sigma$ elasticity of substitution is positive, as expected. However, similar to Albulescu and Pépin (2016), we report a low value for $\sigma$ (0.20), which shows that the leu and the euro are rather complements than substitutes.

The parameters $\hat{\upsilon}_1$ and $\hat{\upsilon}_2$ are significant, underlining that the euro liquidity service compared to that of the leu $(1-\delta(t,\varepsilon_t))/\delta(t,\varepsilon_t)$ considerably diminished. Equation (7) allows to derive an estimate of this ratio date after date (Figure 3).

Figure 3. Estimation of the $(1-\delta(t,\varepsilon_t))/\delta(t,\varepsilon_t)$ ratio

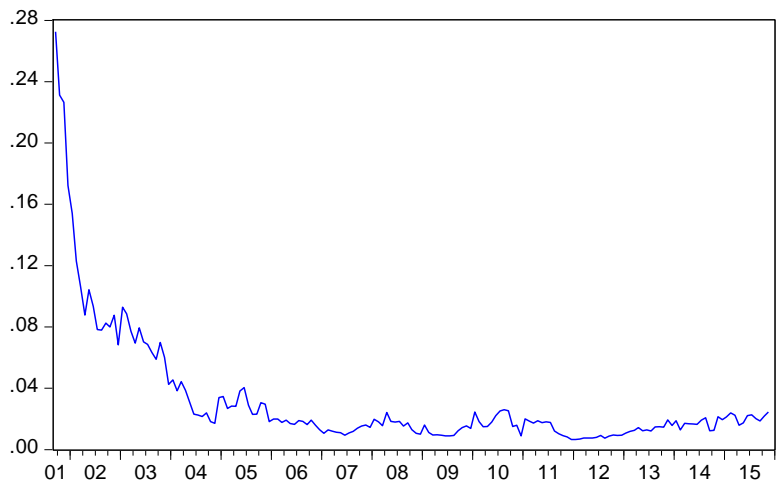

A simple examination of this figure highlights Romanians' loss of interest for the euro. If at the end of 2001 the degree of liquidity of the euro was about 0.25 (indicating that four euros produce the same liquidity service as one leu), it has continuously declined until 2004, with a stabilization around a level of about 0.02. The complementary role of the euro became marginal nowadays.

**Conclusions**

We build a generalized money demand micro-founded model which allows a change in the relative liquidity degree of the euro against the Romanian leu. The long-run money demand is influenced by the interest rate spread, a drop in this spread generating the loss of interest for the euro.

**Acknowledgements**

This work was supported by a grant of the Romanian National Authority for Scientific Research and Innovation, CNCS – UEFISCDI, project number PN-II-RU-TE-2014-4-1760.